\documentclass[intlimits,twoside,a4paper]{article}
\usepackage{graphicx}
\usepackage{wrapfig}
\usepackage[T2A]{fontenc}
\usepackage[cp1251]{inputenc}

\usepackage{cmpj2}

\issue{2014}{17}{3}{33803}
\doinumber{10.5488/CMP.17.33803}
\title[Equilibrium currents in a Corbino graphene ring]%
	{Equilibrium currents in a Corbino graphene ring}

\author[A. L\'opez, N. Bol\'\i var,  E. Medina, B. Berche]
	{A. L\'opez\refaddr{label1}, N. Bol\'\i var\refaddr{label1,label2,label3},
	E. Medina\refaddr{label1,label2,label3},
	 B. Berche\refaddr{label1,label2}}

\addresses{
	\addr{label1} Centro de F\'\i sica,
	Instituto Venezolano de Investigaciones Cient\'\i ficas,
	Apartado 21827, Caracas 1020A, Venezuela
	\addr{label2} Statistical Physics Group,
	P2M Department, Institut Jean Lamour,
        UMR CNRS 7198, BP 70239
        F-54506 Vand\oe uvre les Nancy Cedex, France	
	\addr{label3} Departamento de F\'\i sica,
	Universidad Central de Venezuela,
	Caracas, Venezuela}

\date{Received June 12, 2014}
\begin{document}

\maketitle
\begin{abstract}
We address the description of a graphene Corbino disk in the context of a tight binding approach that
includes both kinetic and Rashba spin-orbit coupling due to an external out-of-plane electric field. Persistent
equilibrium currents are induced by an external magnetic field breaking time reversal symmetry.
By direct diagonalization, we compute the spectrum and focus on the dispersion near the $K$ points at the
Fermi level. The dispersion keenly reproduces that of
a continuum model in spite of the complexity of the boundary conditions. We validate
the assumptions of the continuum model in terms of predominant zig-zag boundaries conditions and weak
sub-band coupling. The wave functions displaying the lowest transverse modes are obtained, showing
the predominance of edge states with charge density at the zig-zag edges. The persistent
charge currents, nevertheless, do not follow the traditional argument of current cancellation
from levels below the Fermi level, and thus they depart in the tight-binding  from those found in the
continuum model.



\end{abstract}

\def\be{\begin{equation}}
\def\ee{\end{equation}}
\def\beq{\begin{eqnarray}}
\def\eeq{\end{eqnarray}}
\def\bpm{\begin{pmatrix}}
\def\epm{\end{pmatrix}}
\newcommand{\<}{\langle}
\renewcommand{\>}{\rangle}
\newcommand{\x}{{\boldsymbol x}}
\newcommand{\y}{{\boldsymbol y}}
\renewcommand{\k}{{\boldsymbol k}}
\newcommand{\q}{{\boldsymbol q}}
\def\bmu{{\boldsymbol \mu}}
\def\bsigma{{\boldsymbol\sigma}}
\def\bdelta{{\boldsymbol\delta}}
\def\bdeltasmall{\footnotesize{\boldsymbol\delta}}
\def\btau{{\boldsymbol\tau}}
\def\boldsymbol#1{\mbox{\boldmath $#1$}}
\newcommand{\bfa}{{\cal W}\!\!\!\!\!\!{\cal W}}
\def\nbOne{\hbox{{\bf 1$\!\!$l}}}
\def\nbM{\hbox{{\bf I\hskip-.8mm M}}_{2\times 2}}
\def\vac#1{{\bf#1}}
\def\bnabla{{\pmb{\nabla}}}

\section{Introduction}
\label{Intro}
Graphene has been a source of recent interest because it is a stable two dimensional material, with
linear dispersion around the Fermi level, and chiral edges states  exhibiting a variety of exciting and tuneable properties~\cite{BookGraphene}. Another reason
for the interest is that in the vicinity of the Dirac cones, many exact results can be obtained and the
tight-binding (TB) approach is perfectly suited, differing very little from first principles calculations.

In the long-wavelength limit, expansion of the kinetic energy around  the Dirac points leads to a set of two Dirac
Hamiltonians
\be
H_0=-\ri\hbar v_{\mathrm{F}}(\tau_z\otimes\sigma_x\partial_x+\nbOne_\tau\otimes \sigma_y\partial_y),
\ee
where $\tau_z$ is the Pauli matrix acting on the valley index which labels the Dirac points
$\vac K=\frac{4\pi}{3a}(1,0)$ and $\vac K'=-\vac K$ and $\sigma_i$ are the Pauli matrices which encode for the two sublattices $\alpha = A,B$. When the Rashba spin-orbit (RSO) interaction, due to an external out-of-plane electric field, is taken into account, the additional term
coupled to the electron spin $\vac s$ appears~\cite{KM05}, namely
\be
H=H_0\otimes\nbOne_s
+\lambda_{\rm R}(\tau_z\otimes\sigma_x\otimes s_y-\nbOne_\tau\otimes\sigma_y\otimes s_x).
\ee
Written in a Corbino ring geometry of average radius $R$ which selects only the lowest transverse modes along the ring, the energy levels can be calculated as~\cite{BMB14},
\be
E_{m,\lambda}^{\kappa,\delta}=\frac\kappa 2\sqrt{
(\hbar v_{\mathrm{F}}/R)^2(1+4m^2)+8\lambda_{\rm R}^2-4\delta
\sqrt{\left[ (m\hbar v_{\mathrm{F}}/R)^2 +\lambda_{\rm R}^2\right] \left[(\hbar v_{\mathrm{F}}/R)^2 +4\lambda_{\rm R}^2\right] }\,.
}
\ee
The parameters $\kappa=\pm 1$ and $\delta=\pm 1$  refer to the particle-hole
and the SO branch indices, respectively. In order to generate persistent charge currents, time reversal symmetry should be broken, e.g. by a
magnetic field flux $\Phi$ piercing the ring. The energy levels are modified according to the substitution
$m\to m+\Phi/\Phi_0$ (with $\Phi_0=2\pi\hbar/|e|$ the flux quantum) which follows from the ordinary $U(1)$ minimal coupling gauge substitution. The
equilibrium charge currents then follow from the relation
\be
J_{\mathrm{Q}}=-\sum_{{\rm occupied\atop\rm states}}\frac{\partial E}{\partial\Phi}\,.
\label{chargecurrentrelation}
\ee
Such a continuum model is particularly simple and the wave functions satisfying the
topology of the ring can be easily obtained~\cite{BMB14}. Nevertheless, there are a few important assumptions regarding the boundary
conditions of the Corbino ring that must be appropriately addressed: i) The internal and external edges of the ring
are neither zig-zag nor armchair, but continuously change as a function of the radial angle. ii) The transverse modes can
be coupled to the longitudinal ones even for the purely kinetic term, as is the case for armchair ribbons and iii) the spin-orbit
coupling is also a source of coupling to transverse modes, so that a single ground state mode theory could be compromised.

In this work we study the validity of the continuum approximation by treating the Corbino geometry through a
tight-binding Hamiltonian approach, thereby correctly incorporating all boundary conditions and verifying the continuum
model assumptions. The paper is organised as follows: We first discuss the tight-binding Hamiltonian contemplating
both the kinetic term and the Rashba coupling. Due to the geometry of the Corbino ring, it is not straightforward to
address the azimuthal momentum within the tight-binding picture. Thus, we make use of a mapping of azimuthal
momentum to the corresponding gauge field, and we can derive the dispersion as its function. The closing
of the wave functions, a condition that must be satisfied for the delocalised electron case, can then be used to
compute the full charge currents.

\section{The tight-binding Hamiltonian}
\label{SecI}
Through the Tight-Binding approach, one can obtain the dispersion in the whole Brillouin zone (not only the vicinity of the Dirac points), and keep track of the hexagonal structure of the lattice. We consider the model Hamiltonian
\be
H = -t_0\sum_{\langle i,j\rangle} \re^{i\phi_{ij}} c_i^\dagger c_j
  + \ri\lambda_{\rm R}\sum_{\<i,j\>}c_i^\dagger\left({\bf s}\times\hat{\bf d}_{ij}\right)_z c_j
 \label{Eq1}
\ee
consisting of two terms.

First the kinetic energy, here written as a hopping term between nearest neighbours (nn) lattice sites on the lattice $\<i,j\>$,  comprises a phase that encodes the magnetic flux.
To evaluate this  Aharonov-Bohm phase, in the basis where Bravais lattice vectors are given by $\vac a_1=a\hat{\vac x}$ and
$\vac a_2=(1/2)a\hat{\vac x}+(\sqrt 3/2)a\hat{\vac y}$ and with the Landau gauge $\vac A=-By\hat{\vac x}$,
one computes, for each link on the lattice, the contribution
\[
\ri\frac{e}{\hbar}\int_{n\vac a_1+m\vac a_2}^{n\vac a_1+m\vac a_2+\bdelta}\vac A\cdot \rd\vac r
\]
with
$\vac r(\eta)=n\vac a_1+m\vac a_2+\eta\bdelta$, $\bdelta$ being one of the $n,m$ site's neighbors.

The second term corresponds to the Rashba spin-orbit interaction due to an external
gate voltage producing an electric field perpendicular to the graphene sheet and
to the strength of the intrinsic spin-orbit interaction \cite{McDonald,Huertas,Fabian}.
The leading contribution results from nearest neighbour terms.


This Hamiltonian can be obtained as follows. First, in the absence of any spin-orbit interactions, the Hamiltonian only contains kinetic energy,
\be
H_0=H(\vac x,\vac p) = \sum_{\vac x,\vac x'}\sum_{\alpha,\alpha'}
|\vac x,\alpha\>\left(-t_{\vac x,\vac x'}^{\alpha,\alpha'}\right)\<\vac x',\alpha'|,
\label{kinetic}
\ee
where the sum is over the lattice sites $\vac x$  of the Bravais lattice and, in general, different orbital types $\alpha$ ($\alpha = A,B$ in the case of graphene). The hopping parameter  $-t_{\vac x,\vac x'}^{\alpha,\alpha'}=\<\vac x,\alpha|H_0|\vac x',\alpha'\>$ is
assumed homogeneous $(=-t_0)$, and creation and annihilation operators can be defined
$c^\dagger_{\vac x,\alpha}c_{\vac x',\alpha'}=|\vac x,\alpha\>\<\vac x',\alpha'|$
in the localized states basis. We usually take into account only nn sites $\vac x$ and $\vac x'$, like in equation~(\ref{Eq1}). The dominant
contribution to the electron transfer between sites come from the $(pp\pi)$, contributions from the unbonded $|p_z\rangle$ orbitals perpendicular to the graphene plane, so the transfer is between orbitals of the same type, and thus $\alpha=\alpha'$ in  equation~(\ref{kinetic}).

The Rashba Spin-Orbit interaction operating in graphene depends on two microscopic contributions; the coupling of atomic
$|p_z\rangle$ to $|s\rangle$ both through the external electric field and the intrinsic SO coupling. The coupling is thus first order in external electric field and first order in the internal atomic electric field sampled by transport electrons. Thus, both the external electric field and the atomic SO couple the atomic $|s\rangle$ and $|p_z\rangle$, and being combined, generate the lowest order coupling~\cite{McDonald}.

The intrinsic SO coupling generates a spin flipping mechanism that can be understood from the form of the bare Pauli Hamiltonian,
\be
H_{\rm Pauli}= \frac{e \hbar}{4m^2c^2}\vac s\cdot (\vac E_{\mathrm{atomic}}\times\vac p),
\ee
where the electric field involved is that coming from the intrinsic atomic field.  In a material, the denominator with the bare mass of the electron should be replaced by the gap in the band structure. This term may be encoded in the kinetic energy term
of the (non-relativistic) electrons through the minimal coupling to a non-Abelian gauge field
$\vac W^a$~\cite{Mineev,Leurs08,MLB08,BBLM09,BML12}
\be
\vac p\longrightarrow\vac p-g\vac W^a\frac{s_a}{2}= \vac p-\frac{e \hbar}{4mc^2}\vac E\times\vac s,
\ee
with $g=\hbar$, the coupling constant to $\vac W^a$. To this gauge field corresponds an unitary
operator
\be
U(\vac x)=\exp \ri\int^{\vac x}\vac W^a(\vac x'')\frac{s_a}{2}\rd\vac x'',
\ee
and the free Hamiltonian $H_0$ transforms according to $UH_0U^{-1}$ in the presence of RSO
interaction:
\beq
H&=&H(\vac x,\vac p-g\vac W^a\frac{s_a}{2})=UH_0U^{-1}\nonumber\\
& =& \sum_{\vac x,\vac x'}\sum_{\alpha,\alpha'}
|\vac x,\alpha\> \<\vac x,\alpha|
\re^{ \ri\int^{\vac x}\vac W^a(\vac x'')\frac{s_a}{2}\rd\vac x''} H_0 \re^{-\ri\int^{\vac x'}\vac W^a(\vac x'')\frac{s_a}{2}\rd\vac x''}
  |\vac x',\alpha'\>\<\vac x',\alpha'|.
\eeq
For a space-independent gauge field, expanding the exponentials to first order, we get
\be
H=-t_0\sum_{\vac x,\vac x'}\sum_{\alpha,\alpha'}
|\vac x,\alpha\> \<\vac x,\alpha|
(1-i\frac{e }{4mc^2}\left(\vac E\times \vac s\right)\cdot\left(\vac x'-\vac x\right)
  |\vac x',\alpha'\>\<\vac x',\alpha'|.
\ee
Keeping only the leading contribution which comes from nearest-neighbouring lattice sites, the quantity
$(\vac E\times \vac s)\cdot(\vac x'-\vac x)$ reduces to $E(\vac s\times\hat{\vac d}_{nn})_z a/\sqrt 3$,
where $a/\sqrt 3=1.42\ \!$\AA~  is the C$-$C distance and $\hat{\vac d}_{nn}$ is the unit vector between
nearest neighbours $\vac x$ and $\vac x'$.
The second term in equation~(\ref{Eq1}) follows, with the Rashba parameter
$\lambda_{\rm R}=\frac{e  Ea(-t_0)}{4\sqrt3 mc^2}$\,. This value of the SO interaction has been derived using the
coupling in the vacuum when in fact it should be that of electrons within the carbon atoms of graphene. Different
authors have derived this form from a Slater-Koster approach~\cite{McDonald,Huertas,Fabian} and second order perturbation theory, resulting in the
form $\lambda_{\rm R}=e  E z_0\xi/3(sp\sigma)$, where $z_0$ is proportional to the atomic size, $\xi$  is the atomic
SO strength, $(sp\sigma)$ is the Slater-Koster integral for the $s-p$ matrix element along the $\sigma$ C$-$C bond.
This is the way atomic SO strengths translate into conduction electron SO strengths.


\section{TB results on the Corbino disk}
\label{SecII}
A Corbino disk is built from the graphene plane, keeping lattice sites which sit between two concentric circles of
given radii (figure~\ref{fig1}). All dangling bonds (sites of coordination one) are eliminated and no account is made for edge deformations~\cite{Snook} due to the reduction in coordination or surface ondulation. Note that there is a dominance of zig-zag edges versus armchair which is general for disks of any size \cite{Akhmerov,Ostaay}.
\begin{figure}[ht]
\centering
\includegraphics[width=10cm]{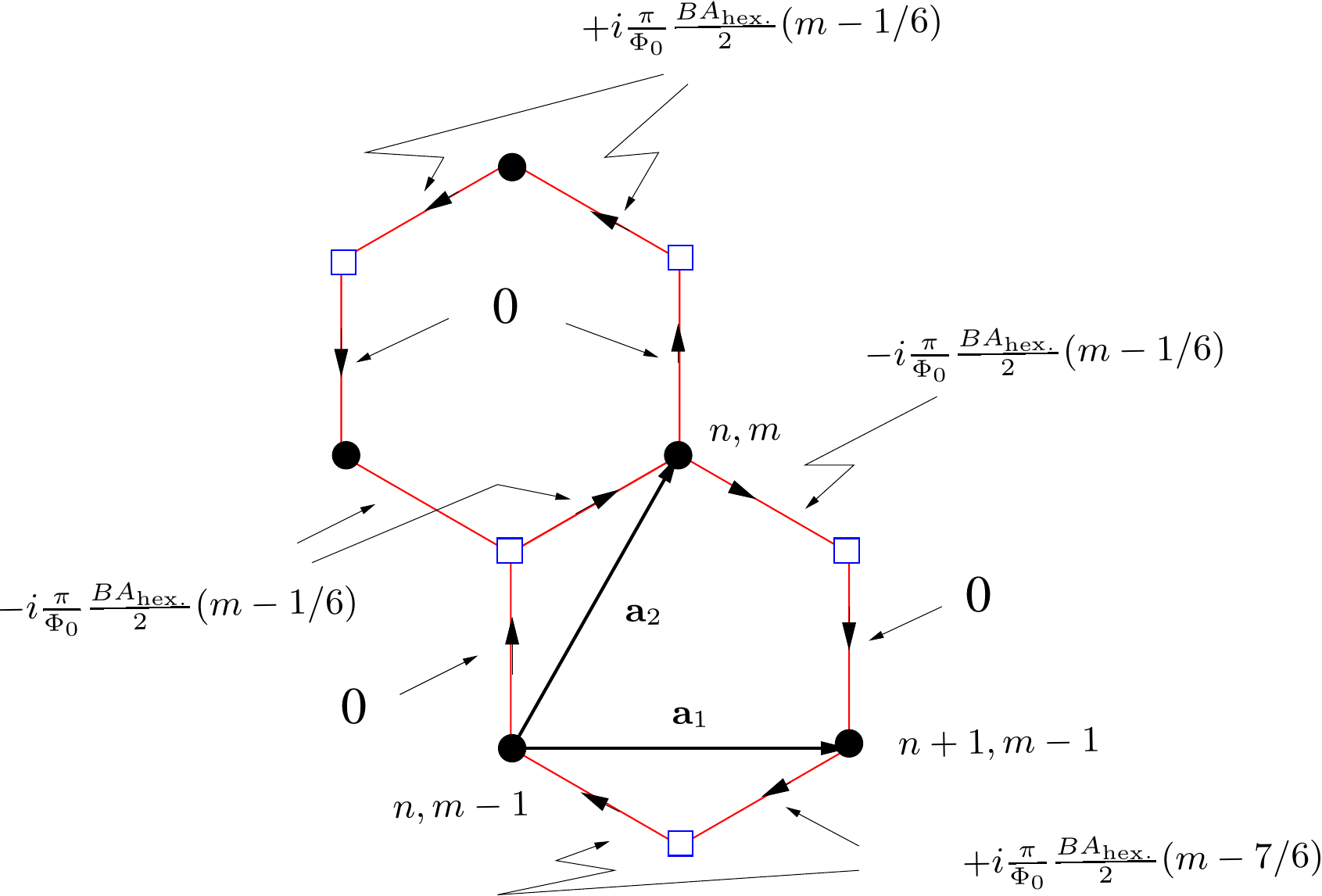}
\caption{(Color online) Aharonov-Bohm phase contributions to the hopping term in the kinetic energy in the TB Hamiltonian.}\label{FigABPhases}
\end{figure}

The TB Hamiltonian of equation~(\ref{Eq1}) is now restricted to the lattice sites forming the disk. The Hamiltonian, restricted to these
sites is diagonalised and eigenvalues and eigenfunctions are derived using Kwant software~\cite{Kwant}.

\subsection{Dispersion relation}
When a magnetic flux is applied across the ring, the magnetic flux being varied may serve as a probe
to scan the Brillouin zone (BZ), as can be seen from the minimal coupling prescription
$\vac p\to\vac p-e \vac A$. The changes in $p$ are symmetric to the changes in $A$. In the
Corbino geometry we have zero radial momentum $p_r$ as there is a fixed transverse state, so the
momentum associated to currents will be purely azimuthal $p_{\varphi}$. We can then use the corresponding component
of the vector potential $A_{\varphi}$ to sweep the azimuthal wave vectors up to a constant.
The wave vectors in the TB approach are measured with respect to the center of the BZ
($\Gamma$ point), but $\Phi$ ($A_{\varphi}$) being changed enables one to approach the Dirac points ($K$ and $K'$ points) by changing
the crystal momentum analog $A_\varphi$ i.e. the magnetic field~\cite{Buttiker,Buttiker2}. At the K points, the valence band and the conduction band cross the Fermi level with linear dispersion in the absence of SO interactions.

\begin{figure}[ht]
\center
  \includegraphics[height=.3\textheight]{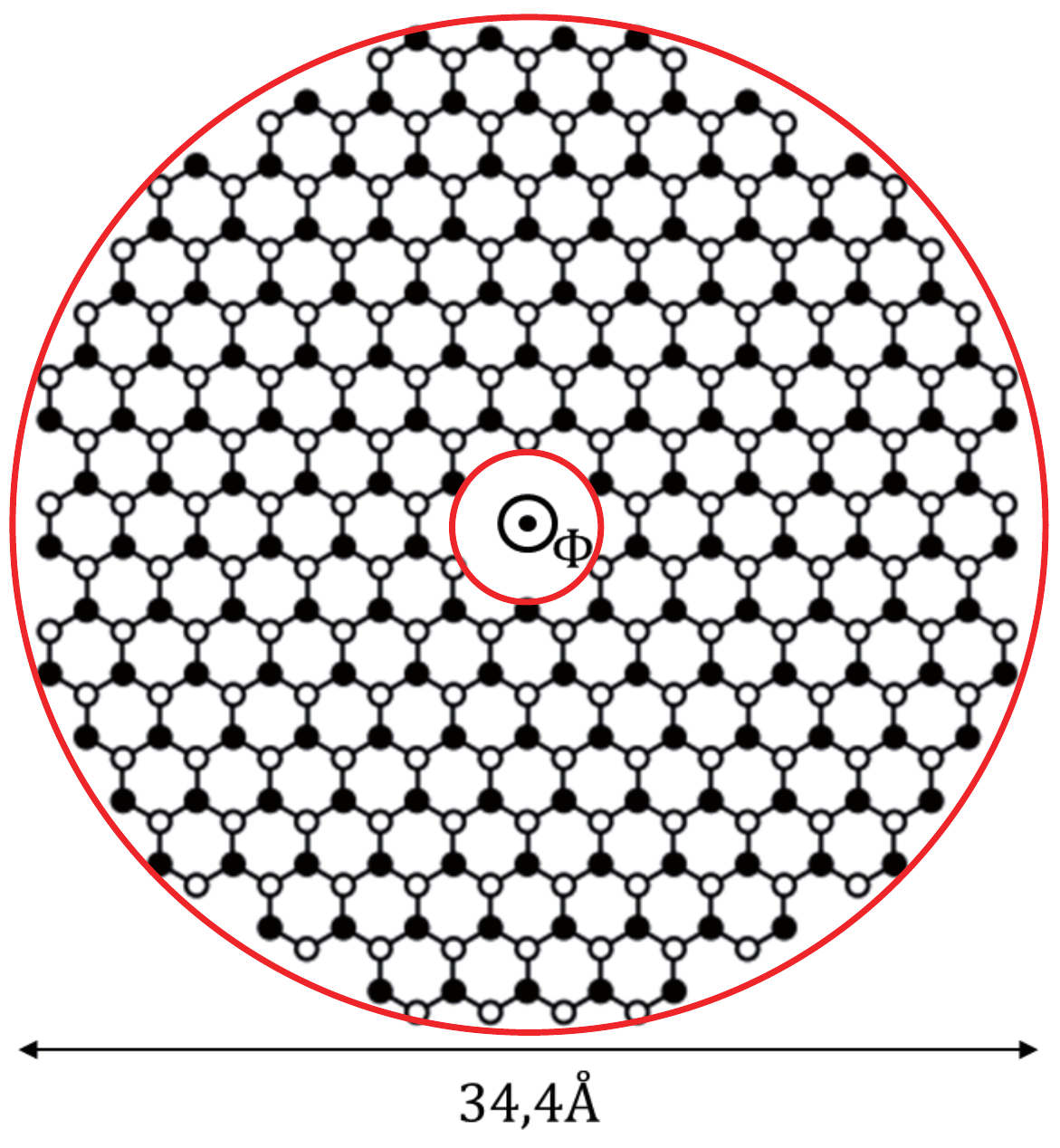}
  \caption{(Color online) Construction of a graphene Corbino disk. We first select the inner and outer radii (the outer radius here is  17.2 \AA, the inner
  radius is 2.5~\AA), then  remove the sites outside the selected area. All dangling bonds or singly coordinated sites are removed. Filled
  and blank sites correspond to carbons on the same sublattice A and B, respectively. A flux $\Phi$ pierces the whole disk.}
\label{fig1}
\end{figure}
\begin{figure}
\center
  \includegraphics[width=7cm]{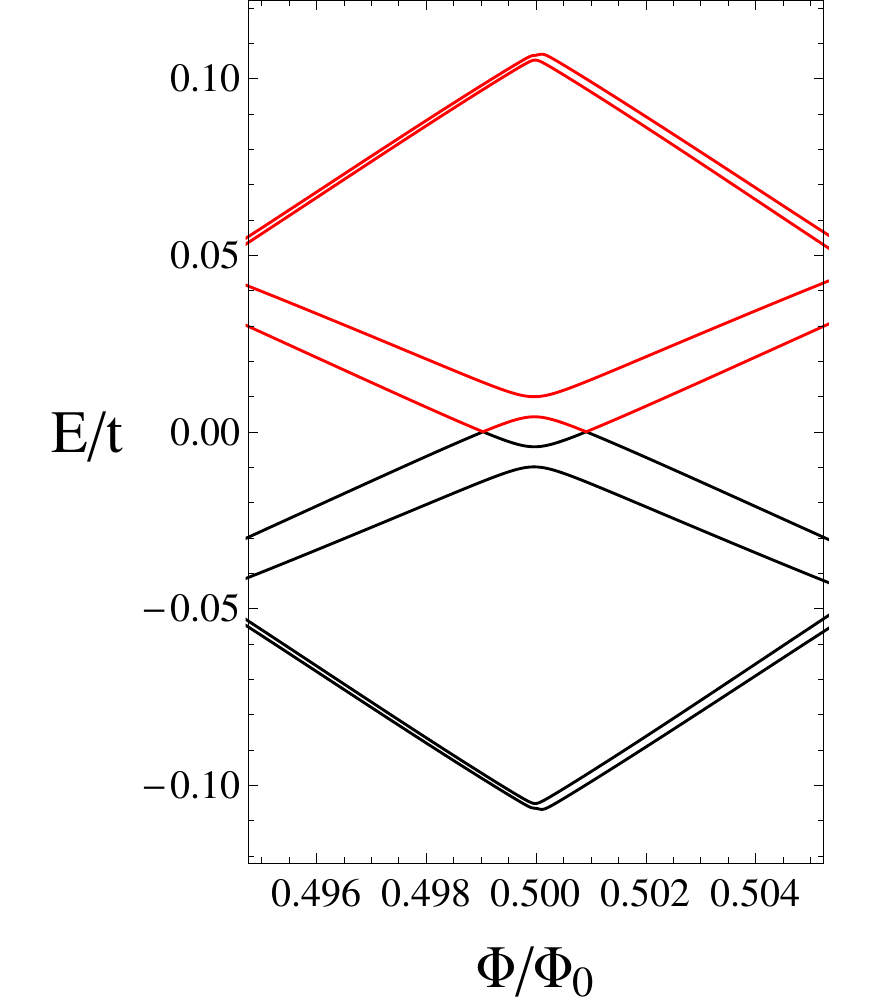}
  \caption{(Color online) The dispersion close to the Fermi energy in the vicinity of $\Phi/\Phi_0=0.5$ and for $\lambda_{\mathrm{R}}=1.2\times 10^{-4}$ a.u., at the K point. We have also depicted
  the subbands closest to the aforementioned structure (not all subbands depicted). The upper subbands being ignored is justified as surmised in the
  continuum approach.}
\label{fig2}
\end{figure}
The dispersion close to the Fermi energy, obtained by following the vector potential, is depicted in figure~\ref{fig2}. We have
also depicted two closest upper and lower sub-bands to assess the energy gap in order to excite those sub-bands. One can see
that this gap is five times the energy scale associated with the width of the energy structure close to the Fermi level. This
energy separation justifies the assumptions made in \cite{BMB14}

\begin{figure}
\includegraphics[height=.365\textheight]{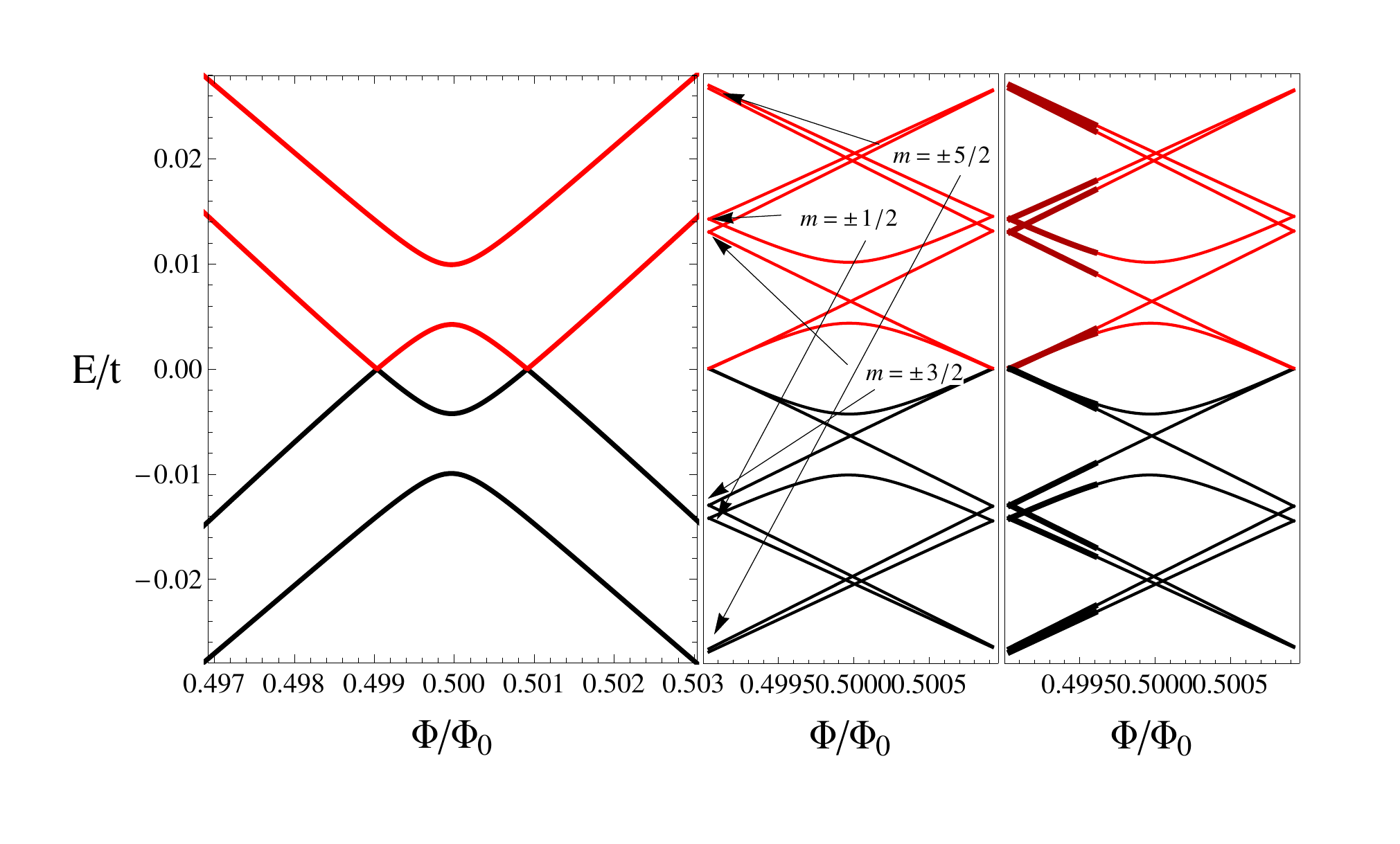}
\caption{(Color online) Dispersion for the Corbino disk in the vicinity of $\Phi/\Phi_0=0.5$. The dispersion on the left panel is that expected
from the continuum model. Once the quantization of the azimuthal momentum defines the quantum number $m$, we can
redraw the excursion of the quantized values as a function of the flux as explained in the text. The right-hand panel depicts the
contribution to the persistent currents at a chosen flux (thick lines).}
\label{threepanels}
\end{figure}

In order to incorporate the quantized values of the energies due to the closure conditions on
the wave function (whose trace we follow with the flux) we recognize that different values for these
quantum numbers follow along this same structure simultaneously, as the field is changed. To reproduce this
evolution, one needs to superpose dephased traces on the same pattern~\cite{BMB14}. This procedure
is illustrated in figure~\ref{threepanels}, where we depict the raw dispersion,
on the left, and the dephased repeated traces on the middle panel, identifying the corresponding
quantum numbers. From the continuum model, the value for the separation between quantised values
is $\Delta\Phi/\Phi_0=\sqrt{2}\sqrt{\epsilon^2+4\lambda_{\mathrm{R}}^2)}$, where $\epsilon=\hbar v_{\mathrm{F}}/R$.
\begin{figure}
\center
\includegraphics[height=.37\textheight]{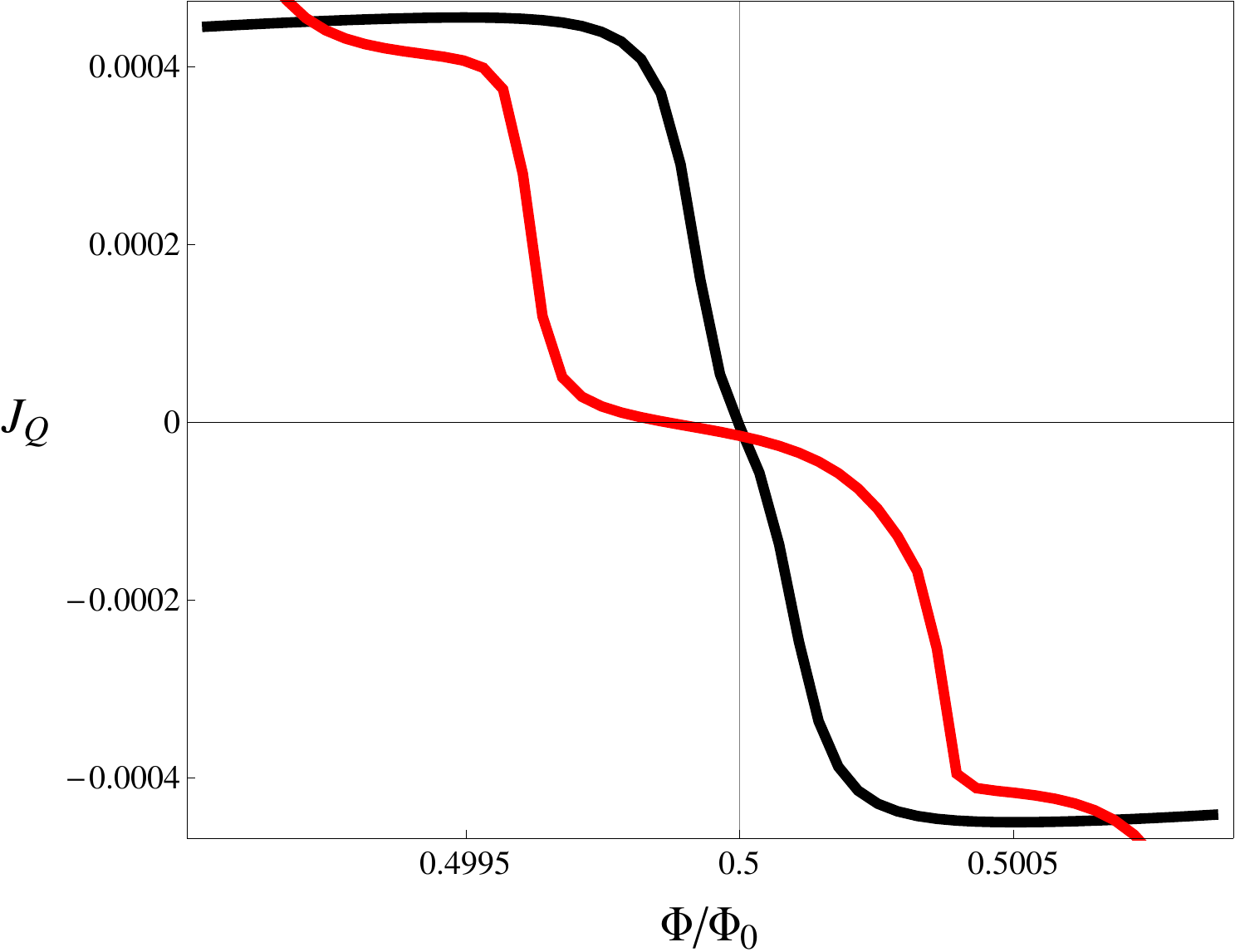}
\caption{(Color online) Persistent charge currents in the vicinity of the K point. The one step curve is purely kinetic Hamiltonian, while the
two-step curve includes the Rashba interaction. The character and magnitude of the current is different
from that of the continuum model due to the finite spectrum and an increasing contribution from the levels deep in the Fermi
sea.}
\label{persistentcurrent}
\end{figure}

\subsection{Charge persistent currents}
Once the energy dependence on the flux is established, we have evidence of delocalised states and thus the possibility
of persistent currents. By referring to figure~\ref{threepanels}, we use equation~(\ref{chargecurrentrelation}) in order to compute the
charge current. While the continuum treatment only contemplated the spectrum depth depicted in the figure (almost identical
to the continuum description), here we have a detailed access to all occupied levels, built from the $2s$ and $2p$ orbitals.

Figure \ref{persistentcurrent} shows the persistent currents derived from the spectrum in the right-hand panel of
figure~\ref{threepanels}. The continuum model by definition has infinite bands and the assumption is that the deep level
bands produce current cancelling contributions. The finite Corbino disk treated here, on the other hand, has a finite
spectrum, and the deep level responds very strongly to the magnetic field. The sum of all contributions then has more
a signature of the deep levels than that of the levels close to the Fermi surface. Whether this makes sense for
finite systems might be a very interesting possibility, since that would mean persistent currents protected from
thermal fluctuations~\cite{BMB14}.

\subsection{Charge density}
\begin{figure}[ht]
\includegraphics[width=0.48\textwidth]{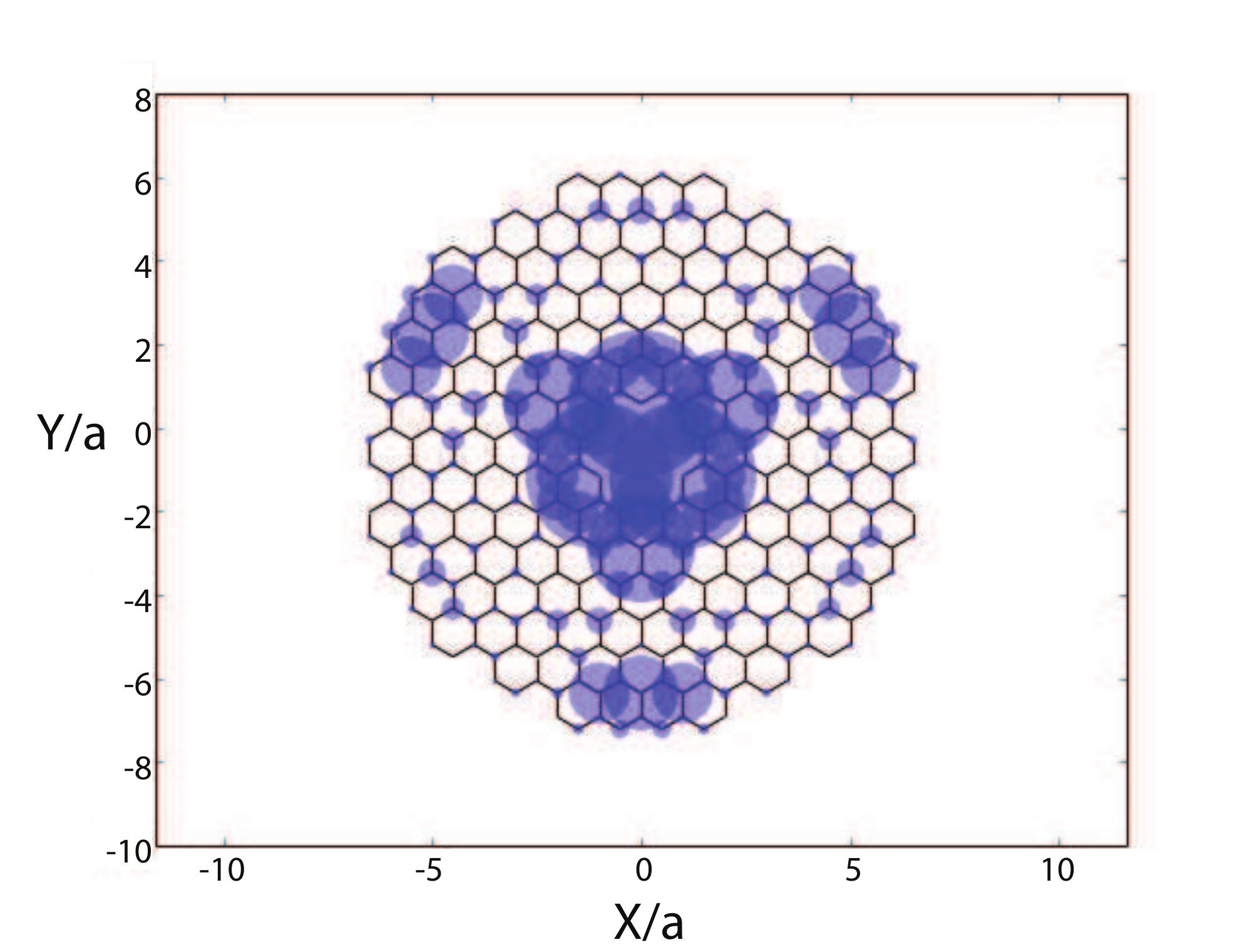}%
\hfill%
\includegraphics[width=0.48\textwidth]{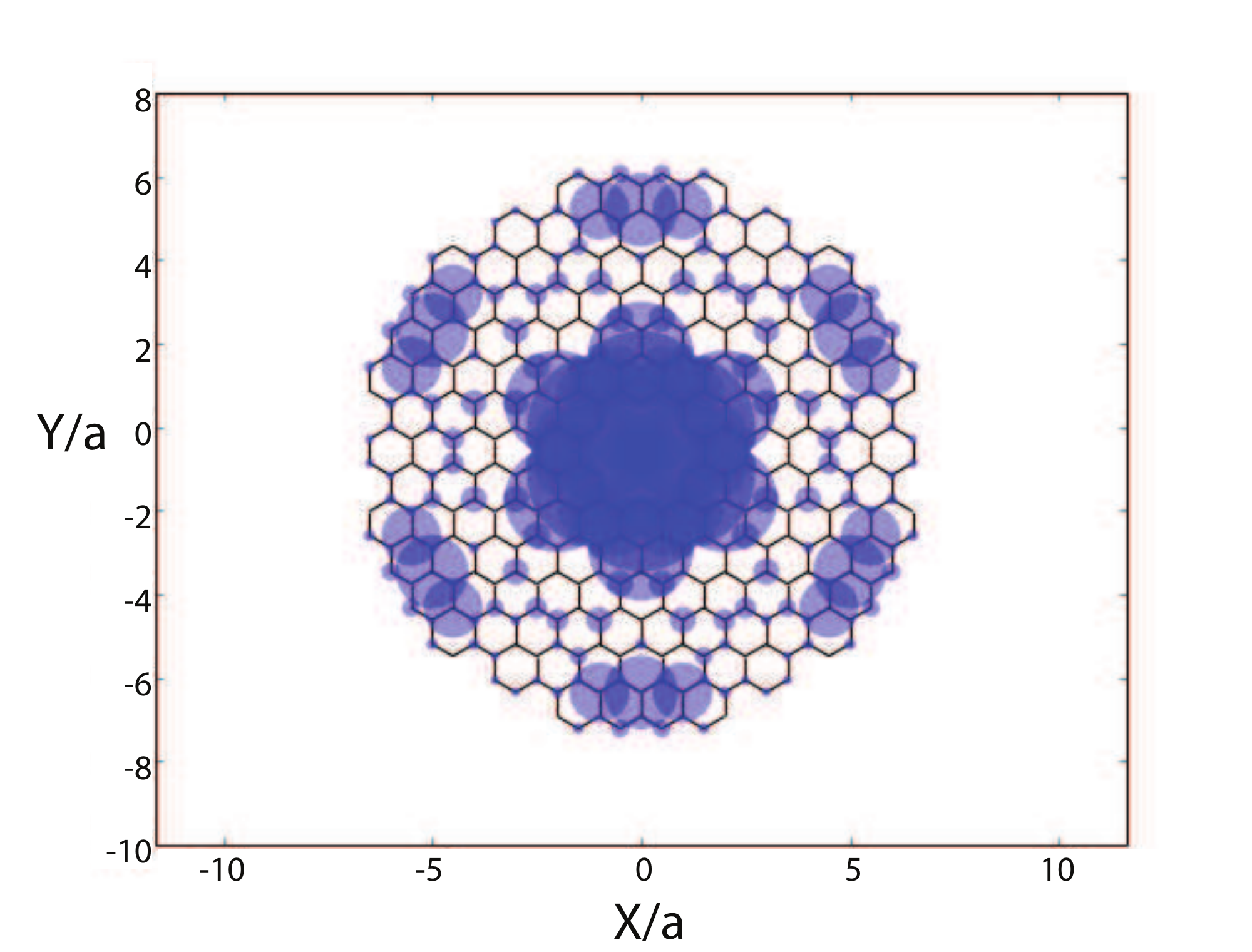}%
  \caption{(Color online) Charge density for the valence spin up band, at the Fermi level. The left-hand panel corresponds to the free case (only kinetic energy), and
  the right-hand panel contemplates Rashba coupling with $\lambda_{\mathrm{R}}=1.2\times 10^{-4}$ a.u. The charge density is concentrated at zig-zag sites in
the exterior of the disk and interior edge.}
\label{fig4}
\end{figure}

In the continuum treatment~\cite{BMB14}, no attention was paid to the details of the lateral confining potential, since the universal features
of the spectrum and the persistent currents, do not depend on it. Nevetheless, the persistent currents are
carried by the edge states, and their configuration is accessible to the TB solution. Figure~\ref{fig4} shows the charge density
on the disk for the ground state, when solely the kinetic term is considered at the $K$ point (left-hand panel).
For the top of the valence band states, the charge piles up on the edges of the disk, almost exclusively on the outer zigzag edges~\cite{AkhmerovGuinea}, with no expected spin distinction. The non-bonding character  of the charge distribution seen for zig-zag edges in nano-ribbons~\cite{Wakabayashi}, does not survive in the Corbino geometry, even in the bulk.  This distribution of charge describes the nature of the ground state transverse mode of the disk.

Once the Rashba SO term is turned on, for sufficiently large values of the coupling, we see a noticeable shift of the charge density
on the edges (figure~\ref{fig4} right-hand panel). While an overwhelming amount of charge is piled up on the zig-zag edges,
there is a modulation in the charge density as compared to the free electron case. Thus, there appears to be a way to tune
the charge distribution on the edges by changing  the electric field perpendicular to the graphene plane. This concentration
of charge is more an indication that additional terms in the Hamiltonian must be contemplated, such as electron-electron interactions
and electron-phonon interactions, which have been associated to edge magnetism~\cite{Wakabayashi} and lattices distortions, respectively.

\section{Conclusions}
We have assessed a tight-binding version of the Corbino disk for graphene in the presence of Rashba SO interactions
and discussed in the light of analytical solutions for the continuum approximation. The assumptions regarding boundary conditions
based on the works in references \cite{Akhmerov,Ostaay} and \cite{Wakabayashi}, where these are predominantly zig-zag and the fact
that a single mode approximation is appropriate, are justified. The full wave functions for the Corbino disk were
drawn for the free and SO coupled cases, where the transverse mode structure revealed both internal and external radius
edge modes.

Regarding the charge current on the disk, there are, nevertheless, qualitative differences arising from the
finite spectrum in the TB approach (while the continuum has an infinite number of states). Current states from a finite set of levels being added up, give rise to contributions from deep in the Fermi sea that change
the currents qualitatively and allow for the possibility of enhanced robustness of persistent current against thermal
fluctuations.

\section*{Acknowledgements} This work was supported by the PICS CNRS--FONACIT programme ``Spin transport and spin manipulations in condensed matter: polarization, spin currents and entanglement'' .
EM and BB are, respectively, grateful to the University of Lorraine and to IVIC for invitations.  Support from the French--German University and the Coll\`ege Doctoral {\it Statistical Physics of Complex Systems} (Leipzig--Nancy--Coventry--Lviv) is also acknowledged (NB). EM acknowledges support from Fundaci\'on POLAR.

\vspace{-3mm}
\ukrainianpart

\title{Рівноважні струми в графеновому кільці Корбіно}

\author{A. Лопес\refaddr{label1}, Н. Болівар\refaddr{label1,label2,label3},
	E. Медіна\refaddr{label1,label2,label3},
	 Б. Берш\refaddr{label1,label2}}

\addresses{
	\addr{label1} Фізичний центр, Венесуельський інститут наукових досліджень,
	Каракас, Венесуела
	\addr{label2} Група статистичної фізики, Інститут Жана Лямура, Вандувр лє Нансі, Франція
	\addr{label3} Фізичний факультет, Центральний університет Венесуели, Каракас, Венесуела}

\makeukrtitle

\begin{abstract}
\tolerance=3000%
Ми звертаємося до опису графенового диска Корбіно в контексті підходу сильного зв'язку, який включає  кінетичний і Рашба спін-орбітовий зв'язок
завдяки позаплощинному електричному полю. Персистентні рівноважні струми індукуються зовнішнім магнітним полем, яке порушує симетрію
часової інверсії. Шляхом прямої діагоналізації, ми обчислюємо спектр і зосереджуємо увагу на дисперсії поблизу
 $K$ точок на рівні Фермі. Дисперсія чітко відтворює дисперсію неперервної моделі, незважаючи на складність граничних умов.
 Ми перевіряємо припущення щодо неперервної моделі в термінах домінантних граничних умов типу зигзагу і слабкого підзонного зв'язку.
Отримано хвильові функції, які демонструють найнижчі поперечні моди і які показують переважання крайових станів із зарядовою густиною на краях зигзагу. Персистентні зарядові струми,
тим не менше,  не слідують традиційному аргументу погашення струмів з рівнів нижче рівня Фермі і тому вони відхиляються
від струмів, знайдених в неперервній моделі.


\end{abstract}

\end{document}